\documentstyle[12pt]{article}
\def\nabstar#1{\nabla\kern-0.5pt\smash{\raise 4.5pt\hbox{$\ast$}}
               \kern-4.5pt_{#1}}

\def\drvstar#1{\partial\kern-0.5pt\smash{\raise 4.5pt\hbox{$\ast$}}
               \kern-5.0pt_{#1}}

\def\newline{\relax\ifhmode\null\hfil\break\else\nonhmodeerr@\newline\fi}
\def\frac#1#2{{#1\over#2}}
\def\text#1{{\hbox{\rm #1}}}
\def\flushpar{{\par \noindent}}

\newcommand{\beq}{\begin{equation}}
\newcommand{\eeq}{\end{equation}}
\newcommand{\bea}{\begin{eqnarray}}
\newcommand{\eea}{\end{eqnarray}}
\def\Id{ \mbox{1\hspace{-1.2mm}I} }
\def\BE{\begin{equation}}
\def\EE{\end{equation}}
\def\BA{\begin{eqnarray}}
\def\EA{\end{eqnarray}}
\def\BAN{\begin{eqnarray*}}
\def\EAN{\end{eqnarray*}}

\def\nn{\nonumber\\}

\def\tr{\mbox{tr}}

\def\det{\mbox{det}}

\def\gm5{\gamma_5}

\input epsf.sty
\newdimen\psfigsize
\def\psfigure#1 #2 #3 #4 #5{
    \begin{figure}[tbh]
      \begin{center}
      \vbox{
        \null\vskip-0.2in\hskip#2
        \epsfxsize=#1
        \epsfbox{#4}
        \vskip -0.3in
        \caption {#5 \label{#3}}
        \vskip 0.0 true in plus 0.3 true in
      }
      \end{center}
   \end{figure}
}
\begin{document}
\thispagestyle{empty}
\begin{flushright}
NTUTH-01-505A \\
June 2001
\end{flushright}
\bigskip\bigskip\bigskip
\vskip 2.5truecm
\begin{center}
{\LARGE {The Index of a Ginsparg-Wilson Dirac Operator}}
\end{center}
\vskip 1.0truecm
\centerline{Ting-Wai Chiu}
\vskip5mm
\centerline{Department of Physics, National Taiwan University}
\centerline{Taipei, Taiwan 106, Taiwan.}
\centerline{\it E-mail : twchiu@phys.ntu.edu.tw}
\vskip 2cm
\bigskip \nopagebreak \begin{abstract}
\noindent

A novel feature of a Ginsparg-Wilson lattice Dirac operator is discussed.
Unlike the Dirac operator for massless fermions in the continuum, 
this lattice Dirac operator does not possess topological zero modes 
for any topologically-nontrivial background gauge fields,
even though it is exponentially-local, doublers-free, and reproduces
correct axial anomaly for topologically-trivial gauge configurations.



\end{abstract}
\vskip 1.5cm

\newpage\setcounter{page}1

In the continuum, the Dirac operator
$ \gamma_\mu ( \partial_\mu + i A_\mu ) $ of massless fermions in a smooth
background gauge field with non-zero topological charge $ Q $ has zero
eigenvalues and the corresponding eigenfunctions are chiral.
The Atiyah-Singer index theorem \cite{Atiyah:1968mp,Atiyah:1971ws} asserts
that the difference of the number of left-handed and right-handed zero modes
is equal to the topological charge of the gauge field configuration :
\bea
n_{-} - n_{+} = Q \ .
\label{eq:AS_thm}
\eea
However, if one attempts to use the lattice \cite{Wilson:1974sk}
to regularize the theory nonperturbatively, then {\it not} every
Ginsparg-Wilson lattice Dirac operator \cite{Ginsparg:1982bj} might possess
topological zero modes\footnote{So far, it has been confirmed that
overlap Dirac operator \cite{Neuberger:1998fp,Narayanan:1995gw} and its
generalization \cite{Fujikawa:2000my}
can possess topological zero modes with index satisfying (\ref{eq:AS_thm}),  
on a finite lattice.} with index satisfying
(\ref{eq:AS_thm}), even though it is exponentially-local, doublers-free, and
reproduces correct axial anomaly for topologically-trivial gauge backgrounds.
As a consequence, a topologically-trivial lattice Dirac operator
might not realize 't Hooft's solution to the $ U(1) $ problem in QCD, nor
other quantities pertaining to the nontrivial gauge sectors. Nevertheless,
from a theoretical viewpoint, it is interesting to realize that one may have
the option to turn off the topological zero modes of a Ginsparg-Wilson
lattice Dirac operator, without affecting its correct behaviors
( axial anomaly, fermion propagator, etc. ) in the topologically-trivial
gauge sector. In this paper, I construct an example of such
Ginsparg-Wilson lattice Dirac operators,
and argue that it does not possess topological zero modes
for any topologically-nontrivial gauge configurations
satisfying a very mild condition, Eq. (\ref{eq:nonzero}).

Consider the lattice Dirac operator
\bea
\label{eq:Dcd}
D = a^{-1} D_c ( \Id + D_c )^{-1}
\eea
with
\bea
\label{eq:Dc}
D_c &=& \sum_{\mu} \gamma_\mu T_{\mu} \ , \hspace{4mm} T_\mu = f t_\mu f \ , \\
\label{eq:f}
f   &=& \left( \frac{1}{\sqrt{t^2 + w^2} + w } \right)^{1/2} \ ,
\hspace{4mm} t^2 = -\sum_{\mu} t_\mu t_\mu \ .
\eea
Here $ \gamma_\mu t_\mu $ is the naive lattice fermion operator
and $ -w $ is the Wilson term with a negative mass $ -1/2 $,
\bea
\label{eq:tmu}
t_\mu (x,y) = \frac{1}{2} [ U_{\mu}(x) \delta_{x+\hat\mu,y}
                       - U_{\mu}^{\dagger}(y) \delta_{x-\hat\mu,y} ] \ ,
\eea
\bea
\gamma_\mu &=& \left( \begin{array}{cc}
                            0                &  \sigma_\mu    \\
                    \sigma_\mu^{\dagger}     &       0
                    \end{array}  \right)  \ ,
\eea
\beq
\sigma_\mu \sigma_\nu^{\dagger} + \sigma_{\nu} \sigma_\mu^{\dagger} =
2 \delta_{\mu \nu} \ ,
\eeq
\bea
\label{eq:w}
w(x,y) =  \frac{1}{2} - \frac{1}{2} \sum_\mu \left[ 2 \delta_{x,y}
                  - U_{\mu}(x) \delta_{x+\hat\mu,y}
                  - U_{\mu}^{\dagger}(y) \delta_{x-\hat\mu,y} \right] \ ,
\eea
where the color and Dirac indices have been suppressed.
Note that the $ D_c $ defined in Eq. (\ref{eq:Dc}) can be regarded as a
symmetrized version of that constructed in Ref. \cite{Chiu:1999hz},
for vector gauge theories.

First, we examine $ D $ in the free fermion limit.
In the momentum space, it can be written as
\bea
\label{eq:Dp}
D(p) = D_0(p) + i \sum_{\mu} \gamma_\mu D_\mu(p) \ ,
\eea
where
\bea
\label{eq:D0p}
D_0(p) &=& \frac{1}{a}
           \left( \frac{ f^4(p) t^2(p) }{ 1 + f^4(p) t^2(p) } \right) \ , \\
\label{eq:Dmu}
D_\mu(p) &=& \frac{1}{a}
        \sin( p_\mu a ) \left( \frac{ f^2(p)}{1+f^4(p) t^2(p)} \right) \ , \\
t^2(p) &=& \sum_{\mu} \sin^2( p_\mu a ) \ , \\
w(p)   &=& \frac{1}{2} -  \sum_{\mu} [ 1 - \cos( p_\mu a ) ] \ ,\\
\label{eq:fp}
f^2(p) &=& \frac{1}{\sqrt{t^2(p)+w^2(p)}+w(p)} \ .
\eea

Now using the relation
\bea
1 + f^4(p) t^2(p) = \frac{ 2 \sqrt{ t^2(p) + w^2(p)} }
                         { \sqrt{t^2(p)+w^2(p)}+w(p) } \ ,
\eea
one can reduce (\ref{eq:D0p}) and (\ref{eq:Dmu}) to
\bea
\label{eq:D0p1}
D_0(p) &=& \frac{1}{2a} \left(1-\frac{w(p)}{\sqrt{t^2(p)+w^2(p)}} \right) \ , \\
\label{eq:Dmu1}
D_\mu(p) &=& \frac{1}{2a} \frac{\sin( p_\mu a )}{\sqrt{t^2(p)+w^2(p)}} \ .
\eea

Evidently, both $ D_0(p) $ and $ D_{\mu}(p) $ are analytic functions for
all $ p $ in the Brillouin zone. ( Note that
$ \sqrt{t^2(p)+w^2(p)} $ is bounded ).
Thus
\bea
\label{eq:Dx}
D(x) = \int \frac{d^4 p}{(2\pi)^4} e^{ i p \cdot x } D(p) \
\eea
is exponentially-local in the position space.
The exponential locality of $ D $ in the free fermion limit immediately
suggests that $ D $ is also exponentially-local for sufficiently
smooth background gauge fields.

In the limit $ a \to 0 $, $ D(p) $ behaves like
\bea
\label{eq:Dp0}
D(p) \sim i  \sum_{\mu} \gamma_\mu p_\mu + O( a p^2 )  \ .
\eea
Thus it has {\it correct continuum behavior}.

The free fermion propagator of (\ref{eq:Dp}) is
\bea
\label{eq:Sp}
D^{-1}(p) = a - i a \sum_{\mu} \gamma_\mu
          \left( \frac{\sin(p_\mu a)}{\sqrt{t^2(p)+w^2(p)}-w(p)} \right) \ ,
\eea
which has a simple pole at $ p=0 $, and does not have any other poles
in the Brillouin zone. Thus it is {\it doublers-free}.

Furthermore, $ D $ is $\gamma_5$-hermitian,
\bea
\label{eq:g5_hermit}
D^{\dagger} = \gm5 D \gm5 \ ,
\eea
and it breaks the chiral symmetry according to the
Ginsparg-Wilson relation \cite{Ginsparg:1982bj}
\bea
\label{eq:gwr}
D \gamma_5 + \gamma_5 D = 2 a D \gamma_5 D \ .
\eea
Thus $ D $ satisfies the necessary requirements for a decent
lattice Dirac operator.

The GW relation (\ref{eq:gwr}) immediately implies
that the fermionic action $ \bar\psi D \psi $ is invariant under
the generalized chiral transformation \cite{Luscher:1998pq}
\bea
\label{eq:ct1}
\psi &\rightarrow& \exp [ i \theta  \gamma_5 ( \Id - a D ) ] \psi, \\
\label{eq:ct2}
\bar\psi &\rightarrow& \bar\psi \exp [ i \theta (\Id- a D ) \gamma_5 ],
\eea
where $ \theta $ is a global parameter.
Consequently, the axial anomaly, $ \tr[ a \gm5 D(x,x) ] $, can be deduced
from the change of fermion integration measure under the exact chiral
transformation (\ref{eq:ct1})-(\ref{eq:ct2}),
and its sum over all sites is equal to the index of $ D $, which is a
well-defined integer \cite{Hasenfratz:1998ri,Luscher:1998pq}
\bea
\label{eq:index_rel}
\mbox{index}(D) = n_{-} - n_{+} = \sum_x \tr[ a \gm5 D(x,x) ] \ ,
\eea
where the trace "tr" runs over the Dirac and color space.
However, the index relation (\ref{eq:index_rel}) does {\it not}
necessarily imply that $ D $ can possess topological zero modes
with the index satisfying the Atiyah-Singer index theorem (\ref{eq:AS_thm}).
In fact, the GW Dirac operator (\ref{eq:Dcd}) always gives
\bea
\label{eq:top_trivial}
n_{+} = n_{-} = \sum_{x} \tr[ a \gm5 D(x,x) ]=0 \ ,
\eea
for any topologically-nontrivial gauge background, even though $ D $
is exponentially-local, doublers-free, $\gamma_5$-hermitian, and
has correct continuum behavior.

The argument is as follows.

From (\ref{eq:g5_hermit}) and (\ref{eq:gwr}), we have
\bea
\label{eq:normal}
D^{\dagger} + D = 2 a D^{\dagger} D = 2 a D D^{\dagger} \ .
\eea
Thus $ D $ is normal and $\gm5$-hermitian. Then the eigenvalues
of $ D $ are either real or in complex conjugate pairs.
Each real eigenmode has a definite chirality,
but each complex eigenmode has zero chirality.
Further, the sum of the chirality of all real
eigenmodes is zero ( chirality sum rule ) \cite{Chiu:1998bh,Chiu:2000tr}.
Now the eigenvalues of $ D $ (\ref{eq:Dcd})
fall on a circle in the complex plane, with center $ ( (2a)^{-1}, 0 ) $
on the real axis, and radius of length $ (2a)^{-1} $. Then the
chirality sum rule reads
\bea
\label{eq:chi_sum_rule}
n_{+} - n_{-} + N_{+} - N_{-} = 0 \ ,
\eea
where $ n_{+} ( n_{-} ) $ denotes the number of zero modes of
positive ( negative ) chirality, and $ N_{+} ( N_{-} ) $
the number of nonzero ( eigenvalue $ a^{-1} $ ) real eigenmodes
of positive ( negative ) chirality.

The chirality sum rule (\ref{eq:chi_sum_rule}) asserts that each
topological zero mode must be accompanied by a nonzero real eigenmode
with opposite chirality, and vice versa. ( Note that both topological zero
modes and their corresponding nonzero real eigenmodes are {\it robust}
under local fluctuations of the gauge background, thus one can easily
distinguish them from those trivial zero and nonzero real eigenmodes which
are unstable under local fluctuations of the background ).

It follows that if $ D $ cannot have any nonzero real eigenmodes in
topologically nontrivial gauge backgrounds, then $ D $
cannot possess any topological zero modes.

From (\ref{eq:Dcd}), any zero mode of $ D $
is also a zero mode of $ D_c $, and vice versa.
However, a nonzero real ( eigenvalue $ a^{-1} $ ) eigenmode of $ D $
corresponds to a pole ( singularity ) in the spectrum of $ D_c $, since
\bea
\label{eq:DcD}
D_c = D ( \Id - a D )^{-1} \ ,
\eea
which is the inverse tranform of (\ref{eq:Dcd}).

Therefore, if the spectrum of $ D_c $ does {\it not} contain any
poles ( singularities ) for a topologically-nontrivial gauge background,
then $ D $ {\it cannot} have any nonzero real eigenmodes,
thus {\it no} topological zero modes.

Now we consider topologically-nontrivial gauge configurations
satisfying the condition
\bea
\label{eq:nonzero}
\det ( \sqrt{t^2 + w^2 } + w ) \ne 0 \ .
\eea
Then $ f $ exists, and $ D_c $
(\ref{eq:Dc}) is well-defined ( without any poles ). It follows that
$ D $ (\ref{eq:Dcd}) cannot have topological zero modes
for any topologically-nontrivial gauge configurations
satisfying (\ref{eq:nonzero}).

It should be emphasized that we have
{\it not} found any {\it robust} nontrivial gauge configuration violating
(\ref{eq:nonzero}), on a finite lattice. Thus, it is likely that
the measure of the nontrivial gauge configurations {\it not} satisfying
(\ref{eq:nonzero}) is {\it zero}.

From (\ref{eq:top_trivial}), the topological triviality of
$ D $ (\ref{eq:Dcd}) implies that it
cannot reproduce correct axial anomaly for topologically-nontrivial
backgrounds. Nevertheless, since $ D $ is exponentially-local, doublers-free
and has correct continuum behavior, these conditions are sufficient to ensure
that it reproduces continuum axial anomaly for topologically-trivial
gauge backgrounds \cite{Chiu:2001ja}.
Further, its exact chiral symmetry guarantees that it
is void of $ O(a) $ artifacts, and is not plagued by the notorious problems
( e.g., additive mass renormalization, mixings between operators in
different chiral representations ) which occur to the
Wilson-Dirac lattice fermion operator.
Therefore it is interesting to investigate to what extent this
GW Dirac operator can provide better chiral properties than the
Wilson-Dirac operator, especially in lattice QCD.
Moreover, it is interesting to compare the
physical observables measured by this GW Dirac operator to those
by the overlap Dirac operator, to understand what role is played by
the topological zero modes.

Finally, it is instructive to unveil the role of the hermitian operator
$ f $ in $ D_c $ (\ref{eq:Dc}).
In the free fermion limit, $ f(p) \simeq 1 + O(a^2 p^2 ) $ for $ p \simeq 0 $.
Thus it retains the physical mode of the naive lattice fermion operator
$ \gamma_\mu t_\mu $. On the other hand, at each one of the $ ( 2^d - 1 ) $
corners of the Brillouin zone ( BZ ), $ f(p) \simeq \infty $ such that
$ f(p) t_\mu(p) f(p) \simeq \infty $.
Thus it decouples all doublers of $ \gamma_\mu t_\mu $, even at {\it finite}
lattice spacing. However the singularities of $ D_c(p) $ at $ ( 2^d - 1 ) $
corners of BZ also render it {\it non-analytic}.
Nevertheless, they do not affect the analyticity of $ D(p) $
since they are cancelled ( from the numerator and denominator )
in the formula (\ref{eq:Dcd}).
Now if $ f(p) $ is analytic for all $ p $
except at $ ( 2^d - 1 ) $ corners of BZ, then $ D(p) $
[ eqs. (\ref{eq:D0p})-(\ref{eq:Dmu}) ] is analytic for all $ p $ in BZ.
Consequently, $ D(x) $ (\ref{eq:Dx}) is exponentially-local.

In general, the basic requirements for the hermitian operator $ f $
in $ D_c $ (\ref{eq:Dc}) are : \\
{\bf (i)}  $ f(p) \simeq 1 + O(a^2 p^2)$ for $ p \simeq 0 $. \\
{\bf (ii)} At each one of the $ ( 2^d - 1 ) $ corners of the Brillouin zone,
$ f(p) \simeq \infty $ such that $ f(p) t_\mu(p) f(p) \simeq \infty $. \\
{\bf (iii)} $ f(p) $ is analytic for all $ p $ except at $ ( 2^d - 1 ) $
corners of the Brillouin zone.

From this viewpoint, we can infer that if $ f $ (\ref{eq:f}) is
replaced by its square ( or $ f^{\alpha} $, $ \alpha > 1/2 $ ),
\BAN
\label{eq:f2}
f = \frac{1}{\sqrt{t^2 + w^2} + w } \ ,
\EAN
then the resulting $ D(p) $ is also analytic, doublers-free, and has
correct continuum behavior.

Presumably, one might also construct entirely new examples of $ f $
satisfying above basic requirements {\bf (i)-(iii)}.

In passing, we note that the example defined in
Eqs. (\ref{eq:Dcd})-(\ref{eq:f}) is only a special case ( $ c = 1/2 $ )
of the following GW Dirac operator
\bea
\label{eq:Dtw}
D &=& a^{-1} D_c ( \Id + r D_c )^{-1} \ ,  \hspace{4mm} r = \frac{1}{2c} \ , \\
\label{eq:Dcc}
D_c &=& \sum_{\mu} \gamma_\mu f t_\mu f \ , \\
f &=& \left( \frac{2c}{\sqrt{t^2 + w^2} + w } \right)^{1/2} \ ,
\hspace{4mm} t^2 = -\sum_{\mu} t_\mu t_\mu \ ,  \nn
w(x,y) &=&  c  - \frac{1}{2} \sum_\mu \left[ 2 \delta_{x,y}
                  - U_{\mu}(x) \delta_{x+\hat\mu,y}
                  - U_{\mu}^{\dagger}(y) \delta_{x-\hat\mu,y} \right], \ \
  0 < c < 2 \ . \nonumber
\eea
In the free fermion limit, (\ref{eq:Dtw}) gives
\BAN
D(p) = \frac{c}{a} \left( 1-\frac{w(p)}{\sqrt{t^2(p)+w^2(p)}} +
i \sum_\mu \gamma_\mu \frac{\sin( p_\mu a )}{\sqrt{t^2(p)+w^2(p)}} \right) \ ,
\EAN
which is analytic, doublers-free, and has correct continuum behavior.
Further, there are many viable forms of $ D_c $.
For example, a variant of $ D_c $ is
\BAN
D_c &=& \sum_{\mu} \gamma_\mu f t_\mu f^{\dagger} \ , \\
f &=& \left[ (\sqrt{t^2 + w^2}-w) \frac{2c}{t^2} \right]^{1/2} \ ,
\EAN
which agrees with $ D_c $ (\ref{eq:Dcc}) in the free fermion limit.
In this case the condition for gauge configurations (\ref{eq:nonzero})
should be replaced by
\BAN
 \det(t^2) \ne 0 \ .
\EAN
These lattice Dirac operators all satisfy the necessary requirements for
a decent lattice Dirac operator, namely, exponential-locality, doublers-free,
correct continuum behavior, $\gamma_5$-hermiticity and the Ginsparg-Wilson
relation. However, they do {\it not} possess topological zero modes.

For some years, it has been taken for granted that if a Ginsparg-Wilson
lattice Dirac operator has correct axial anomaly for the trivial gauge
sector, then it must also reproduce continuum axial anomaly for the
nontrivial sectors. However, the lattice Dirac operator (\ref{eq:Dcd})
provides a counterexample, and suggests that this common conception may
{\it not} be justified.

In general, given a topologically-proper
lattice Dirac operator, it can be
transformed into a topologically-trivial lattice Dirac operator which
is identical to the topologically-proper one in the free fermion
limit. On the other hand, given a topologically-trivial GW
Dirac operator, it remains an
interesting question how to transform it into a topologically-proper one.

\eject

\bigskip
\bigskip
\flushpar
{\bf Acknowledgement }
\bigskip

\noindent

I would like to thank David Adams and Herbert Neuberger for their
feedback on the first version of this paper.
This work was supported in part by the National Science Council,
Republic of China, under the grant number NSC89-2112-M002-079.

\bigskip
\bigskip


\end{document}